\documentstyle[12pt]{article}
\def\TL{\hfil$\displaystyle{##}$}
\def\TR{$\displaystyle{{}##}$\hfil}

\def\TT{\hbox{##}}
\def\seqalign#1#2{\vcenter{\openup1\jot
  \halign{\strut #1\cr #2 \cr}}}

\def\comment#1{}
\def\fixit#1{}

\def\overleftrightarrow#1{\vbox{\ialign{##\crcr
     $\leftrightarrow$\crcr\noalign{\kern-0pt\nointerlineskip}
     $\hfil\displaystyle{#1}\hfil$\crcr}}}

\def\lsim{\mathrel{\mathstrut\smash{\ooalign{\raise2.5pt\hbox{$<$}\cr\lower2.5pt\hbox{$\sim$}}}}}
\def\gsim{\mathrel{\mathstrut\smash{\ooalign{\raise2.5pt\hbox{$>$}\cr\lower2.5pt\hbox{$\sim$}}}}}

\def\sqr#1#2{{\vcenter{\vbox{\hrule height.#2pt
         \hbox{\vrule width.#2pt height#1pt \kern#1pt
            \vrule width.#2pt}
         \hrule height.#2pt}}}}

\def\href#1#2{#2}  

\def\lbldef#1#2{\expandafter\gdef\csname #1\endcsname {#2}}
\def\eqn#1#2{\lbldef{#1}{(\ref{#1})}%
\begin{equation} #2 \label{#1} \end{equation}}
\def\eqalign#1{\vcenter{\openup1\jot
    \halign{\strut\span\TL & \span\TR\cr #1 \cr
   }}}

\def\comment#1{  \begin{raggedright}{\tt [#1]}\end{raggedright}}
\def\fixit#1{}

\def\comment#1{  \begin{raggedright}{\tt [#1]}\end{raggedright}}
\def\fixit#1{}

\def\ie{{\it i.e.}}
\def\eg{{\it e.g.}}

\def\al{{\alpha}}

\def\ra{\rightarrow}
\def\CN{{\cal{N}}}
\begin{document}
\baselineskip=15.5pt
\pagestyle{plain}
\setcounter{page}{1}


\begin{titlepage}

\begin{flushright}
CLNS-01/1746 \\
PUPT-1996 \\ 
hep-th/0107111
\end{flushright} 
\vfil
 
\begin{center}
{\Large Hot Little String Correlators: A View From Supergravity}
\end{center}

\vspace{0.5cm}
\begin{center}
{\large K.~Narayan$^{a,}$\footnote{narayan@mail.lns.cornell.edu}
and Mukund Rangamani$^{b,}$\footnote{mukund@feynman.princeton.edu}}
\end{center}

$$\seqalign{\span\TL & \span\TT}{
^a & Newman Laboratory, Cornell University, Ithaca, NY 14853.
 \cr\noalign{\vskip1\jot}
^b & Joseph Henry Laboratories, Princeton University, Princeton,
NJ 08544.
}$$
\vspace{1cm}

\begin{center}
{\large Abstract}
\end{center}
We study the propagation of a massless minimally coupled scalar in the 
near horizon geometry of non-extremal NS5-branes. Using the holographic 
principle for dilatonic backgrounds we compute the two-point function 
of an operator in Little String Theory at the Hagedorn temperature. We
then comment on relations with correlation functions in two dimensional
string theory.

\vspace{6cm}
\begin{flushleft}
July 2001
\end{flushleft}

\end{titlepage} 
\section{Introduction}

The worldvolume description of $N$ coincident Neveu-Schwarz five-branes is
a Poincare invariant non-gravitational field theory in $5+1$ dimensions, 
referred to as the Little String Theory (LST). This theory is defined
\cite{seiberg} to be the decoupled theory on the world-volume in the limit 
of vanishing string coupling $g_s \rightarrow 0$ (for a review see 
\cite{aharony}). Even though the string coupling is taken to zero, the 
theory retains nontrivial dynamics and is nonlocal \cite{kapustin}. 
The decoupling limit for NS5-branes ensures that these theories are 
intrinsically strongly coupled, for they have no continuous dimensionless
coupling constant, instead being characterized by a single integer $N$, 
the number of NS5-branes. In the infrared limit with sixteen supersymmetries,
these theories flow to the $(2,0)$ superconformal field theory in $5+1$ 
dimensions (in the Type IIA description) and to IR free $5+1$ SYM with 
$(1,1)$ supersymmetry (in the Type IIB description).

Despite having successfully defined decoupled non-local theories on the 
world-volume of NS5-branes, it has proven quite hard to extract substantial
information about the dynamics of the theory. Efforts in this direction 
include formulation of a light-cone description \cite{abs}, which has 
proved useful in extracting the chiral operator spectrum and the 
formulation of a holographic dual in terms of string theories in linear 
dilaton backgrounds \cite{abks}. The holographic description has been used
to extract some correlation functions \cite{minsei} and has been extended
to more general theories with fewer supersymmetries \cite{gkpelc, gka, gkb}.

LSTs can also be defined at finite temperature. A useful technique to
study the dynamics of such a system is to construct a holographic dual to 
the LST at a given temperature, which typically is a black hole geometry. 
At finite temperature, the decoupling limit for NS5-branes is defined by 
taking the asymptotic value of the string coupling $g_s$ to zero while 
keeping fixed the string scale, $l_s = \sqrt{\al'}$. By tuning the number 
$N$ of coincident five-branes and the dimensionless energy density above 
extremality $\mu$ on the branes, the holographic dual supergravity becomes 
a valid description for the parameter range $\mu \gg N \gg 1$ \cite{ms}. 
It turns out however that this simple holographic dual exists only for a 
particular value of the temperature which happens to coincide with the 
Hagedorn temperature of the little strings. It was shown in \cite{ms} that
LST at the Hagedorn temperature is holographically dual to string theory on
the CHS tube \cite{chs} capped off by a horizon (the two-dimensional Euclidean
cigar geometry), along with an $SU(2)$ WZW model with level prescribed by
the number of NS5-branes and a free CFT for the longitudinal directions 
of the brane.

Correlation functions in LST at zero temperature were calculated in 
\cite{minsei} by studying the propagation of a minimal scalar in the 
supergravity background of N coincident IIA NS5-branes. Since the geometry 
in the Type IIA description is the linear dilaton background, the growing 
of the dilaton as we descend down the CHS tube invalidates the use of 
ten dimensional supergravity all through. Instead what one does is to use
the M-theory lift of the configuration: start with a stack of coincident 
M5-branes evenly arrayed along the 11th direction ${\bf S}^1$. 
In the limit of a large number of M5-branes, one can trust the 11-dimensional 
supergravity results. The full geometry has the linear dilaton throat 
interpolating between asymptotically flat space and the near horizon
geometry of a stack of M5-branes {\it i.e.}, $AdS_7 \times {\bf S}^4$.
In the decoupling limit the asymptotic flat region is pushed away to infinity.
One prescribes boundary conditions for the scalar at some hypersurface 
$ r = \Lambda$, and computes the Euclidean action as a function of the 
boundary values, in the general spirit of holography. The resulting 
correlation function for the operator that couples to the scalar
was found to reduce consistently to that of the $(2,0)$ superconformal
theory in the IR. One interesting aspect of the analysis was that the 
absorption cross-section was found to be non-zero in the decoupling limit 
for modes with energies larger than $m_s/\sqrt{N}$ (which in the large $N$ 
limit is much lower than the string scale), the mass gap in the throat. 

In what follows, we compute the two-point function for a minimally coupled
scalar in LST at the Hagedorn temperature, by exploiting the 
aforementioned holographic description. 
We consider the Euclidean 
geometry with a cut-off imposed up the tube far from the horizon. This 
cut-off is equivalent to an ultraviolet cut-off in the dual field theory. We 
then define a Dirichlet boundary value for the minimal scalar, demanding 
that the solution go to a constant value at finite distances up the tube 
as the cutoff is removed. Imposing the constraint of regularity at the 
horizon, the scalar field can be uniquely solved in the
bulk. We then evaluate the Euclidean action on the bulk scalar solution 
to the Dirichlet problem, which reduces to a cut-off dependent boundary
term. The two-point function of the operator dual to the bulk scalar is 
then computed from the leading finite part after dropping a divergent 
piece. This is similar to the corresponding computations of two-point 
functions in $AdS$ backgrounds \cite{gkp9802, witten9802} and their 
noncommutative counterparts \cite{malru}. 

The basic result of the paper is the two point function of an operator 
that couples to a massless minimal scalar propagating in the bulk
geometry. It agrees with the two-point function (for the momentum sector) 
obtained from a CFT analysis in double scaled LST \cite{gka, gkb}. Further,
the two-point function contains a factor which is very similar to the 
``external leg factors'' that appear in the context of two dimensional 
string theory \cite{igor, ginsmoore}. There are simple poles on the real 
momentum axis, from these external leg factors.

This paper is organized as follows. In section 2, we describe the 
supergravity background holographically dual to LST at the Hagedorn 
temperature. In section 3, we describe the computation of the two-point
function. In section 4, we elaborate on properties of the two point 
function thus obtained and finally in section 5, we discuss certain 
features that are analogous to correlation functions calculated in
two dimensional string theory. Appendix A outlines the
computation of the reflection coefficient in the Lorentzian geometry.

\section{The Holographic Dual}

The metric of the non-extremal NS5-brane in the string frame \cite{ms}
is given as (suppressing the NS-NS $B_{\mu \nu}$ field that the NS5-branes
couple to) \footnote{Note that this metric is related to that appearing
in \cite{hs91}, \cite{gs9202} by a simple coordinate transformation.}

\eqn{undstrmet}{\eqalign{
ds_{str} ^2 & = -f(r) dt^2 + dy_5^2 + A(r) \left( {dr^2 \over f(r)} +
 r^2 d\Omega_3^2
\right) \cr 
e^{2 \Phi} &= g_s^2 A(r) \cr
f(r) &= 1 - {r_0^2 \over r^2} \cr
A(r) &= 1 + {N l_s^2\over r^2} \equiv 1 + {r_5^2 \over r^2}
}}
The $dy_5^2$ corresponds to the flat spatial directions along the 
5-branes, while the $d\Omega_3^2$ corresponds to the 3-sphere part of 
the transverse geometry. $r_0$ is a non-extremality parameter, the location
of the outer horizon being $r = r_0$. The geometry transverse to the 
5-branes is a long tube which opens up into the asymptotic flat space 
region with the horizon at the other end. In the extremal limit, \ie\ 
$r_0 = 0$, the above geometry factorizes into ${\bf R}^{5,1} \times 
{\bf R}_{\phi} \times {\bf S}^3_N$, where ${\bf R}_{\phi}$ stemming from the 
radial part of the metric represents the linear dilaton tube. The dilaton
grows towards the horizon.

The decoupling limit \cite{ms} is defined as a double-scaling limit, 
scaling the asymptotic value of the string coupling and the horizon 
radius to zero simultaneously, keeping fixed the energy density above 
extremality in string units \ie, 

\eqn{decolim}{
g_s \ra 0, \;\;\;\; r_0 \ra 0, 
\;\;\;\;\; \mu = {r_0^2 \over g_s^2 \ l_s^2} = \rm{fixed}
}

For the purposes of our calculation, we resort to Euclidean space. We 
will thus work in the Wick-rotated version of the metric \undstrmet\ and 
simply drop the $1$ in $A(r)$ in the decoupling limit. 

To analyse this limit it is convenient to introduce a new coordinate
$r = r_0 \cosh \sigma$, which in the scaling limit, gives for the 
Euclidean near-horizon geometry

\eqn{scalim}{\eqalign{
ds_{str}^2 &= \tanh^2 \sigma \;dt^2 + Nl_s^2 \; d\sigma^2 + Nl_s^2 \;
d\Omega_3^2 + dy_5^2 \cr
 e^{2\Phi} &= {N \over \mu \ \cosh^2 \sigma} 
}}

\noindent
The spacetime geometry is smooth in the parameter range $\mu \gg N \gg 1$.
Further in this range, string perturbation theory is also good.
The geometry in this case becomes ${\bf R}^5 \times {\cal M}_{2dbh} 
\times {\bf S}^3_N$. ${\cal M}_{2dbh}$ represents the spacetime corresponding 
to the Euclidean 2D black hole. The 3-sphere is of constant radius 
$\sqrt{N \al'}$. Both string frame and Einstein frame tube lengths, computed
using \scalim, are infinite in the scaling limit. Thus in the decoupling 
limit, the geometry consists of a semi-infinite tube of constant size capped
by the black hole horizon at one end. In other words, the asymptotic flat 
space region is infinitely far away from the horizon.

The Hawking temperature of this spacetime can be calculated by the 
usual methods, \ie, by calculating the periodicity of the temporal part 
of the Wick-rotated metric. In other words, we demand that the 
Euclidean metric does not have a conical singularity at the origin. This 
gives
\eqn{hawktemp}{
T_H = {1 \over 2\pi \sqrt{N \al'}}
}
The Hawking temperature is independent of the horizon size $r_0$ and 
therefore of the energy density $\mu m_s^6$. It only depends on the 
number $N$ of 5-branes. This independence of the temperature on the energy
implies that the entropy is proportional to the energy, giving rise to a
vanishing free energy at leading order in supergravity. Basically the 
thermal ensemble is degenerate, we shall return to this point in the 
discussion.

\section{Scalar propagation}

We want to study the issues related to scattering of minimally coupled 
complex scalar particles in the geometry of the non-extremal NS5-brane. 
The classical action for a massless minimally coupled complex scalar is
given as 

\eqn{scalaraction}{
S = {\kappa  \over g_s^2} \int d^{10} x 
\ \sqrt{G_E} \ \ |\partial \phi|^2
}

\noindent
We can think of a massless minimally coupled scalar as being a 
longitudinally polarized mode of the graviton. The normalization 
factor $\kappa$ will be fixed at a later stage.
 
For an $s$-wave scalar with no momentum along the longitudinal directions
and $ \phi(t,r,y) = e^{i \omega t + i k \cdot y} \phi(r)$, the wave equation
reduces to

\eqn{euscalar}{
{f(r) \over r^3} \; \partial_r \left( r^3\;  f(r) 
\partial_r\phi \right) \; - \; 
A(r) \left( \omega^2 + k^2 f(r) \right) \phi(r) =  0
}

\noindent
As mentioned previously, we work in the Wick rotated version of 
the non-extremal metric in \undstrmet\ and since we will be interested 
in setting boundary conditions in the CHS tube, we shall drop the $1$ from
the harmonic function $A(r)$.  
Incidentally the very same equation was obtained in the analysis of 
\cite{msgbdy} and \cite{km}. These authors were interested in the 
computation of greybody factors for scattering off 5-dimensional 
black holes.  We shall use the results of the latter as they work in the 
regime more suited to our analysis. 

What we are really after is the two-point function of an operator that 
couples to the scalar $\phi$ in the form given by 

\eqn{interaction}{ 
S_{int} = \int d^6p \; \left( \phi^*(p,\Lambda) {\cal{O}}(p) + \phi(p,\Lambda) 
{\cal{O}}(p)^* \right),
}

\noindent
where $ r = \Lambda$ is the UV cut-off in the dual field theory. 

The two-point function of the operator ${\cal{O}}$ in momentum space is 
defined to be 
$ \ \langle {\cal{O}}(p) {\cal{O}}^*(-p') \rangle = \Pi(p) \delta(p-p')$. 
Given a solution to the equations of motion implied by
\scalaraction\ we evaluate the Euclidean 
action on the solution. It is imperative that we pick the right boundary 
conditions for the scalar field. The boundary conditions 
we wish to impose on the solution are the following:

\begin{itemize}
\item{The solution has to be regular at the horizon {\it i.e.}, at the 
tip of the cigar in the Euclidean case.}
\item{We will impose a cut-off at some distance along the linear-dilaton
tube, say at $r =\Lambda$. We require that the solution be constant at this 
surface as we take the cut-off to infinity.}
\end{itemize}

The boundary condition at the horizon is physically reasonable. The boundary 
condition at the cut-off surface is chosen in accord with the intuition 
gained in the analysis of massive scalars in $AdS$, {\it cf.}, \cite{kwitten}, 
and a very analogous scenario seen in the case of the holographic 
dual to the non-commutative gauge theories \cite{malru}.
The point is the following: As we shall see explicitly below, demanding 
regularity at the horizon leads to a solution that has an asymptotic 
behaviour with a growing and a damped piece far up the CHS tube. In the 
general spirit of holography one would like to impose a Dirichlet boundary 
condition at the cut-off surface. A natural way to implement the same in the 
present set-up is to have the scalar field to remain constant at large radii
as we take the cut-off to infinity. This is a renormalization prescription 
that turns out to depend on the momenta of the scalar, reflecting the 
non-local nature of the theory \cite{peetpol,minsei}. 

Evaluation of the Euclidean action \scalaraction\ on a solution with the 
aforementioned boundary conditions leads to a boundary term

\eqn{actonsol}{\eqalign{
S &= {\kappa V_5 \Omega_3 \over 2 g_s^2 }\;
\left[ r^3 f(r) \phi^*(r) \partial_r \phi(r) \right] \mid_{r_0}^\Lambda 
+ \ {\rm c.c.} \cr
  &= {\kappa V_5 \Omega_3  \over 2 g_s^2} \; f(\Lambda) \Lambda^3 \; 
\phi^*(\Lambda) 
\; \left[ \partial_r \phi(r) \right] \mid_{r=\Lambda} + \ {\rm c.c.}
}}

\noindent
This can be evaluated and after discarding a piece that diverges as the 
cut-off is taken to infinity we are left with a well-defined two-point 
function for the operator ${\cal{O}}$.

The equation \euscalar\ is exactly solvable in terms of Hypergeometric 
functions, {\it cf.}, \cite{km}. Substituting $ z = f(r)$ we obtain

\eqn{kmhg}{
z {d \over dz} z {d \over dz} \phi(z) + { r_5^2 \over 4 (1-z)^2 }(\omega^2 
+ k^2 z) \phi(z)= 0.
}

\noindent
which is solved by 
\eqn{hypgeosoln}{\eqalign{
\phi(z) & = A z^{\alpha_{+}} ( 1 - z )^{\beta} F(\alpha_+ + \beta, \alpha_+ + \beta;
1 + 2 \alpha_{+}; z) \cr
& \;\;\; + B z^{\alpha_{-}} ( 1 - z )^{\beta} F(\alpha_- + \beta, \alpha_- + \beta;
1 + 2 \alpha_{-}; z).
}}

\noindent
for constants $A$ and $B$, where, 
\eqn{pars}{\eqalign{
\alpha_{\pm} &= \pm {\omega r_5 \over 2} = \pm{\sqrt{s} \over 2 } \cr
\beta &= {1 \pm \sqrt{ 1 + r_5^2 (\omega^2 + k^2)} \over 2} 
= {1 \pm \sqrt{1 + s + k^2 r_5^2} \over 2},
}}
defining $s = \omega^2 r_5^2 = \omega^2 N \al'$, the energy squared in 
units of the mass gap. Note that in the Euclidean black hole background, 
$\omega$ is quantized in units of $1/\sqrt{N \al'}$ which is related to 
the Hawking temperature $T_H$ (see sec.2). Thus $s$ takes values of 
integer squared.

The first boundary condition causes us to reject the solution with the 
negative root for $\alpha$ (For $z \rightarrow 0$, which is the location
of the horizon, the hypergeometric function goes to 1). We shall 
henceforth drop the subscript in $\alpha_+$ and also choose the smaller 
root for $\beta$ (this choice turns out to be arbitrary since choosing 
the larger root does not change our results). 

The solution $ \phi(z) = z^{\alpha} (1-z)^{\beta} F(\alpha + \beta, \alpha + 
\beta, 2 \alpha + 1; z)$ asymptotes at large $r$ ($z \ra 1$) to

\eqn{philarge}{\eqalign{
\phi(r) & \sim \CN(p) \left[ 
\left({r_0 \over r}\right)^{2\beta} 
+ B(p) \left({r_0 \over r}\right)^{2(1-\beta)} \right]
\cr
& \sim {r_0 \over r} \CN(p) \left[ 
\left({r_0 \over r}\right)^{-\sqrt{1+s+k^2r_5^2}}
+ B(p) \left({r_0 \over r}\right)^{\sqrt{1+s+k^2r_5^2}} \right],
}}

\noindent
where

\eqn{defbs}{\eqalign{
B(p) & = {\Gamma(2 \beta - 1) \over \Gamma(1 - 2\beta)}
\; {\Gamma^2(1 + \alpha - \beta)  \over \Gamma^2(\alpha + \beta)} 
\cr & = 
{\Gamma(-\sqrt{1 + s + k^2 r_5^2}) \over \Gamma(\sqrt{1 + s + k^2 r_5^2})} 
\ {\Gamma^2({1 + \sqrt{s} \over 2} + {\sqrt{1 + s + k^2 r_5^2} \over 2})
\over \Gamma^2({1 + \sqrt{s} \over 2} - {\sqrt{1 + s + k^2 r_5^2} \over 2})}.
}}

\noindent
Demanding that we have 
\eqn{bndrycondn}{
\phi(\Lambda, p) = \left({r_0 \over 
\Lambda}\right)^{1 - \sqrt{1+s+k^2r_5^2}} \phi_0(p)
= \left({r_0 \over \Lambda}\right)^{2\beta} \phi_0(p),
}
we fix 
\eqn{normalization}{
\CN(p) \sim { 1 \over 1 + B(p) \left({r_0 \over 
\Lambda}\right)^{2 \sqrt{1+s+k^2r_5^2}}}.
}

We are now well-positioned to evaluate the Euclidean action \actonsol. 
Using $f(\Lambda) \sim 1$ and 

\eqn{derphi}{
\partial_r \phi  \sim \CN(p) \left[- {2\beta
\over r_0}\left({r \over r_0}\right)^{-2\beta - 1} 
 - B(p) {2(1-\beta) \over r_0} \left({r \over r_0}
\right)^{-2(1-\beta) - 1}
\right] }

\noindent
implying,

\eqn{eval}{
\phi^*(r) \partial_r \phi(r) |_\Lambda 
\sim |\CN(p)|^2 \ \left[- {2\beta \over r_0}
\left({\Lambda \over r_0}\right)^{-4\beta - 1} 
- 2 B(p) {r_0^2 \over \Lambda^3} \right].
}

\noindent
The first term in \actonsol\ diverges and we drop it, the second term 
will give (note $\CN(p) \sim 1)$)

\eqn{eaction}{
S =  -2 \kappa V_5 \Omega_3 \;{r_0^2 \over g_s^2} \;B(p) 
\;\phi_0(p) \phi_0(-p)
= -2 \kappa V \Omega_3 l_s^2 \mu \; B(p) \;\phi_0(p) \phi_0(-p).
}

Differentiating w.r.t. the sources $\phi_0$, we obtain, apart from overall 
normalization factors, the two-point function

\eqn{twoptfn}{
\langle {\cal{O}}(p) {\cal{O}}^*(-p') \rangle \sim  
{\Gamma(-\sqrt{1 + s + k^2 r_5^2}) \over \Gamma(\sqrt{1 + s + k^2 r_5^2})} 
\ {\Gamma^2({1 + \sqrt{s} \over 2} + {\sqrt{1 + s + k^2 r_5^2} \over 2})
\over \Gamma^2({1 + \sqrt{s} \over 2} - {\sqrt{1 + s + k^2 r_5^2} \over 2})}
\ \delta(p-p').
}

\noindent
We can choose $\kappa$ and the normalization of the operator 
such that \twoptfn\ is the correctly normalized two-point function of 
the operator ${\cal O}$ and in the following we shall do so accordingly. 
So henceforth we shall take the right hand side of \twoptfn\ to be the 
appropriately normalized two-point function of the operator ${\cal O}$ that 
couples to a massless minimal scalar.

\section{Properties of the two-point function}

We have calculated the two-point function $\Pi(p)$ as a function 
of momentum $p = (\omega, k)$. Note that since the theory is at 
finite temperature, we have broken Lorentz invariance and thus the spatial
momenta $k$ do not appear on the same footing as the energy $\omega$. 
$\omega$ is quantized in units of $1/r_5 = 1/{\sqrt{N \al'}}$, the 
periodicity of the time variable in the black hole background, which is 
related to the Hawking temperature (see sec.2). This implies that 
$s = \nu^2, \ \nu = 0,1,2,\ldots$ From \twoptfn, we have 
\eqn{Pip}{
\Pi(p) = 
{\Gamma(-\sqrt{1 + s + k^2 r_5^2}) \over \Gamma(\sqrt{1 + s + k^2 r_5^2})} 
\ {\Gamma^2({1 + \sqrt{s} \over 2} + {\sqrt{1 + s + k^2 r_5^2} \over 2})
\over \Gamma^2({1 + \sqrt{s} \over 2} - {\sqrt{1 + s + k^2 r_5^2} \over 2})}
}
Note that with the assignments \footnote{We thank D.~Kutasov for pointing
this out to us.}
\eqn{gkmatch}{
\sqrt{1 + s + k^2 r_5^2} = 2j + 1, \qquad
\sqrt{s} = \nu = -2m = 2{\bar m},
}
$\Pi(p)$ coincides with the two-point function given in eqn.(3.6) of 
\cite{gkb} in the large $N$ limit (with small $j$), 
\eqn{Pijm}{
\Pi(j,m) = {\Gamma(-2j-1) \over \Gamma(2j+1)} \ {\Gamma^2(j - m + 1) \over
\Gamma^2(-j - m)}
}
upto an overall constant, which has to do with the normalization of the 
operators. Here $(m, {\bar m})$ parametrize momentum and winding around 
the cigar via
\eqn{mbarm}{
m = {1\over 2} (-\varrho + w N), \qquad 
{\bar m} = {1\over 2} (\varrho + w N).
}
Note that $\varrho = -p$ in eqn.(2.3) of \cite{gkb}. Since our computation
corresponds to momentum modes on the cigar geometry, the winding number 
$w = 0$ and we have $m = -{\bar m}$. We do not see the first ratio of 
$\Gamma$-functions in eqn.(3.6) of \cite{gkb} since our semiclassical 
analysis holds only for small $j/N$.

$\Pi(p)$ has a relatively simple analytic structure. There is a tower of 
simple poles from the first ratio of $\Gamma$-functions in \Pip\ for 
\eqn{polesA}{
p^2 r_5^2 = s + k^2 r_5^2 = \nu^2 + k^2 r_5^2 = n^2 - 1,
\qquad n = 1, 2, \ldots
}
from the $\Gamma$-function in the numerator \footnote{Note that the 
possible pole at $n=0$ is absent because it cancels with a similar singular
factor from the denominator $\Gamma$-function.}. Since the $\Gamma$-function
has no branch cuts, the only branch cuts in $\Pi(p)$ arise from the square
roots in the arguments of the $\Gamma$-functions. The absence of branch cuts
(apart from the square roots) in \Pip\ might suggest that within the 
approximations of our calculation, only single-particle states are visible 
from the couplings to the bulk observables we use, while multi-particle 
states are not (see also \cite{gka}).

Of course, not all of the above are poles of the two-point function, since
the squared $\Gamma$-function in the denominator too has poles, which 
compensates. For instance, setting $s = \nu^2 = 0$, the poles are at 
$1 + k^2 r_5^2 = n^2$. The denominator has $\Gamma^2({1-n\over 2})$ which 
has double poles when $ n $ is an odd integer. So basically one half of the 
poles actually end up being zeros instead of the two-point function. 
Nonetheless, there is a tower of poles on the real axis, given by \polesA\
with $n$ being a nonzero even integer, all of whose residues have the same 
sign. These poles become more and more evenly spaced in $|p|$, as $p$ grows
large. The poles do not coincide with the locations of the quantized 
frequencies.

Since on-shell bulk fields give rise to off-shell correlation functions in
the dual field theory, $\nu$ and $k$ are independent variables above. They
are not related by a six dimensional mass-shell condition.

Note that $\Pi(p)$ is really the renormalized two-point function. The cutoff
$\Lambda$ appears in our boundary condition \bndrycondn, for large $p$, as 
\eqn{lambdalargep}{
\Lambda^{ |p| \sqrt{N \al'} } = e^{ |p| \sqrt{N \al'} \ {\rm log} \Lambda }.
}

Consider, for \eg, performing a ``wave-function renormalization''
in our interaction action \interaction. If we rescale $\phi$ by this 
exponential factor thereby rescaling the operator ${\cal O}$ by the inverse 
factor, the unrenormalized two-point function rescales by the square of this
factor, \ie\ $e^{2 |p| \sqrt{N \al'} \ {\rm log} \Lambda}$, which grows 
exponentially for large $p$, corroborating with \cite{peetpol}, 
\cite{minsei}, \cite{kapustin}.

It would be interesting to perhaps incorporate our results in modelling
these nonlocal nongravitational field theories with a view to obtaining 
insight into string interactions in these systems (see, for \eg\, \cite{br}
for discussions on extensivity in these systems).

\section{Discussion}

To shed some light on the poles we see in the above two-point function, 
let us recall a similar feature \footnote{We thank I.~Klebanov for 
valuable discussions on these issues.} that is seen in the analysis of 
2-dimensional gravity coupled to $c=1$ matter \cite{igor, ginsmoore}. 
Correlation functions of tachyon vertex operators in that model
have poles arising from 
the so called ``external leg factors''. These leg factors
are proportional to ${\Gamma(-|k_i|) \over \Gamma(|k_i|)}$, 
$k_i$ being the momenta of the tachyon. In the limit of large momenta 
the argument of the first ratio of Gamma functions in \twoptfn\ is 
identical to the leg factors. 
We of course have other factors in the correlation function, 
but they are incapable of producing any poles, and as we have seen 
they can at best project out some of the poles. In the $c=1$ story, the poles 
occur at integral momenta and are understood to be indicative of special 
states in the theory occurring only at these momenta
\cite{klebpol91}, \cite{ew91}. They can be thought of
as remnants of the transverse excitations of the string in two spacetime 
dimensions.

To get some insight into these aspects, realize that we have 
constructed momentum states in the asymptotic geometry
of the supergravity background \undstrmet\ and computed the reflection 
coefficient for scattering these states off the background. Since the 
linear dilaton region of the geometry has a mass-gap of 
$ m_{gap} = {1 \over \sqrt{N \alpha'}}$, the argument of the first ratio 
of $\Gamma$-functions in $\Pi(p)$ is the energy of the excitation above 
the mass gap. In the two-dimensional string case the argument of the 
$\Gamma$-function in the external leg factor can be thought of as a 
Liouville energy, since the vertex operator satisfying the mass-shell 
condition is  $e^{i kX} e^{(-2 + |k|) \Phi}$, $\Phi$ being the Liouville 
coordinate. By Liouville energy we simply mean the factor multiplying 
$\Phi$ in the exponent, modulo a universal factor, which in the units we 
choose to write the vertex operators happens to be $-2$. 
 
Now consider trying to formulate the physical state conditions for
vertex operators in the case we are studying : the background in string 
frame is ${\bf R}^5 \times {\bf S}^3 \times {\cal M}_{2dbh}$ where 
${\cal M}_{2dbh}$ corresponds to the 2D black hole cigar geometry. 
This geometry asymptotes to ${\bf R}^5 \times {\bf S}^3 \times 
{\bf R}_{\phi} \times {\bf R}_t$, with $R_{\phi}$ corresponding to the 
linear dilaton direction and in addition we have 
$g_s^2 \sim e^{-2 \sigma/\sqrt{N \al'}}$ as can be seen from \scalim\ 
($\sigma$ parametrizes the linear dilaton direction). Consider a graviton 
vertex operator (with no excitation on the ${\bf S}^3$) in the linear 
dilaton region of the geometry, 
\eqn{Vop}{
\xi_{\mu \nu} \psi^\mu {\bar \psi}^\nu e^{i k \cdot X } e^{\alpha \Phi},
}
$\psi^\mu$ being the worldsheet fermions and $\xi_{\mu \nu}$ being the 
polarization tensor. We have dropped the ghost contributions for simplicity
(see eqn.(3.21) of \cite{gka}, for the vertex operator written in the 
$SL(2)/U(1)$ coset language). Then the mass shell condition is 

\eqn{massshell}{
{1\over 2} k^2 - {1\over 2} \alpha (\alpha - Q) = 0 
}

\noindent
where $Q$ is the Liouville charge, which can be 
determined from the asymptotic behaviour of the dilaton (also {\it c.f.}, 
eqn.(2.18) of \cite{ks}) to be $ Q = {2 \over \sqrt{N \alpha'}}$.
From the mass-shell condition \massshell\ we can determine the 
Liouville energy $\alpha$ to be 

\eqn{alphak}{
\alpha = {Q \over 2} \pm \sqrt{k^2 + {Q^2 \over 4}}
}

The first term corresponds to the shift in the Liouville energy arising from
the background while the second term is exactly the same as the arguments
in the first ratio of $\Gamma$-functions in $\Pi(p)$ for 
$Q = 2/\sqrt{N \al'}$.

It is therefore tempting to conjecture that the poles in the two-point 
function \twoptfn\ correspond to special states in LST at the 
Hagedorn temperature, whose vertex operators are of the form given in 
\Vop. More precisely it is likely that the physical state 
conditions for the vertex operators of the above form have extra solutions 
at momenta which satisfy the conditions in \polesA\ in analogy with the 
situation in two dimensional string theory
as was demonstrated in \cite{klebpol91}. It would be interesting 
to analyze the special states in the present case along these lines.

In the analysis of double scaled LST \cite{gka,gkb} there were additional 
poles arising from modes that had winding about the asymptotic circle.
These were understood to correspond in the semi-classical description 
to bound states living near the tip of the cigar \cite{dvv}.
The spectrum of the bound states has been known to correspond exactly with 
the principal discrete series representation of $SL(2,R)$. We however 
have no winding about the asymptotic circle and therefore are not in a position
to reproduce these poles in our analysis.

Little String Theories as we have seen are defined holographically 
at the Hagedorn temperature. At tree level this implies that when thought of 
as a thermal field theory, the thermodynamics is degenerate, since the 
energy and the temperature can be tuned independently. 
A proper treatment of the thermal ensemble would need to ask what changes 
would be induced by quantum fluctuations. It turns out that a one-loop 
analysis of the thermodynamics reveals an instability \cite{ks}:
to be precise, the one-loop specific heat is negative.
It was argued in \cite{mr} that this instability would imply that the 
background geometry \scalim, suffers from a marginal case of 
Gregory Laflamme instability \cite{gl, reall}, 
generalising the ideas of \cite{gubm}.

In light of the instability of the geometry one might be skeptical about 
the nature of the computation presented above. The basic point is that the 
geometry \scalim\ is only marginally unstable (or better, neutrally 
stable). In particular the tree level supergravity analysis ought to reveal
only a massless mode and so the geometry is metastable.
In particular, our results for the two-point function ought to capture 
the leading behaviour.

\section*{Acknowledgements}

We wish to thank O.~Aharony, P.~Argyres, T.~Becher, M.~Berkooz, V.~Hubeny,
M.~Krasnitz, J.~Maldacena, V.~Sahakian, A.~Sen, H.~Verlinde, 
E.~Witten and especially 
I.~Klebanov and D.~Kutasov for extremely fruitful discussions. The work 
of KN was partially supported by NSF-grant PHY95-13717. The work 
of MR was partially supported by NSF-grant PHY-9802484.

\section*{Appendix A : Reflection coefficient in the tube}

In this section, we describe the calculation of the reflection coefficient
for scalar modes propagating in the geometry with energies 
$m_s/\sqrt{N} < \omega << m_s$. This is usually calculated demanding 
ingoing boundary conditions at the horizon (this is equivalent to 
regularity in the interior in the Euclidean calculation). In addition, 
we demand that the solution be constant on a cutoff surface
along the linear dilaton tube, as the cutoff is taken to infinity. The 
solution that results thus is pure ingoing at the horizon and has both
ingoing and outgoing components far from the horizon. The action evaluated
on this solution is used after subtracting off the action evaluated on free
propagation in the tube (this ``free'' action corresponds to the divergent
piece in the Euclidean calculation). This subtracted action is then used 
to compute the reflection coefficient.

\noindent
The wave equation for a massless minimal scalar reduces to
\eqn{scalarM}{
{f(r) \over r^3} \; \partial_r \left( r^3 \  f(r) 
\partial_r\phi \right) \ + \
A(r) \left( \omega^2 + k^2 f(r) \right) \phi(r) =  0
}
Using the substitution $ z = f(r)$, the resulting equation can be solved 
to give
\eqn{hypgeosolnM}{
\phi(z) = z^{\alpha} ( 1 - z )^{\beta} F(\alpha + \beta, \alpha + \beta;
1 + 2 \alpha; z).
}

\noindent
Here we have defined new parameters $\alpha$ and $\beta$ which 
are given as 
\eqn{parsM}{
\alpha = - i {\sqrt{s} \over 2}, \qquad
\beta  = {1 - \sqrt{ 1 - s + k^2 r_5^2} \over 2}
}
where the lower root for $\beta$ has been chosen. The negative root for 
$\alpha$ is chosen since we are interested in an ingoing wave at the 
horizon. The other independent solution of the second order differential
equation above 
corresponds to the outgoing wave solution and is thus discarded.

\noindent
This solution asymptotes at large $r$, \ie\ $z \ra 1$, to (we set $k=0$ 
here for simplicity)
\eqn{philargeM}{\eqalign{
\phi(r) & \sim \CN(s) \left[ 
\left({r_0 \over r}\right)^{2\beta} 
+ B(s) \left({r_0 \over r}\right)^{2(1-\beta)} \right]
\cr
& \sim {r_0 \over r} \CN(s) \left[ 
\left({r_0 \over r}\right)^{-\sqrt{1-s}}
+ B(s) \left({r_0 \over r}\right)^{\sqrt{1-s}} \right].
}}
where
\eqn{defbsM}{
B(s) = {\Gamma(2 \beta - 1) \over \Gamma(1 - 2\beta)}
\; {\Gamma^2(1 + \alpha - \beta)  \over \Gamma^2(\alpha + \beta)}
}

\noindent
Demanding that we have $\phi(\Lambda,\omega) = \left({r_0 \over 
\Lambda}\right)^{1 - \sqrt{1-s}} \phi_0(\omega)$,
we fix the normalization $\CN(s)$.

The reflection coefficient is obtained from the ratio of the ingoing 
and the outgoing wave pieces in the solution far from the horizon, \ie\ 
$B(s)$. Alternatively, calculating the action from the solution thus 
obtained and subtracting a divergent piece corresponding to free motion 
in the tube gives the reflection coefficient $B(s)$ upto other factors.


\end{document}